\documentclass[aps,amsmath,amssymb,prb,twocolumn,showpacs,floatfix]{revtex4}
\usepackage{bm}
\usepackage{epsfig}

\newcommand{\ud}{{\textrm d}}

\newcommand{\bq}{{\bf q}}
\newcommand{\bQ}{{\bf Q}}
\newcommand{\bk}{{\bf k}}
\newcommand{\bqaf}{{\bf q}_{\rm AF}}

\newcommand{\nn}{{\nonumber}}

\newcommand{\coma}{{\it cond-mat}}

\begin{document}

\title{Magnetic fluctuations in n-type high-$T_c$ superconductors reveal breakdown of fermiology}

\author{F. Kr\"uger$^1$, S. D. Wilson$^2$, L. Shan$^3$, S. Li$^2$, Y. Huang$^3$, H.- H. Wen$^3$, S.-C. Zhang$^4$, Pengcheng Dai$^{2,5}$, J. Zaanen$^1$}

\affiliation{$^{1}$Instituut-Lorentz, Universiteit Leiden, P. O. Box 9506, 2300 R A Leiden, The Netherlands\\
$^2$Department of Physics and Astronomy, The University of Tennessee, Knoxville, Tennessee 37996-1200, USA\\
$^3$National Laboratory for Superconductivity, Institute of Physics and National Lab for Condensed Matter Physics, Beijing 100080, China\\
$^4$Department of Physics, Stanford University, Stanford, California 94305-4045, USA\\
$^5$Neutron Scattering Sciences Division, Oak Ridge National Laboratory, Oak Ridge, Tennessee 37831-6393, USA}

\begin{abstract}
By combining experimental measurements of the quasiparticle and dynamical magnetic properties of optimally electron-doped Pr$_{0.88}$LaCe$_{0.12}$CuO$_4$ with theoretical calculations we demonstrate that the conventional fermiology approach cannot possibly account for the magnetic fluctuations in these materials. In particular, we perform tunneling experiments on the very same sample for which a dynamical magnetic resonance has been reported recently and use photoemission data by others on a similar sample to characterize the fermionic quasiparticle excitations in great detail. We subsequently use this information to calculate the magnetic response within the conventional fermiology framework 
as applied in a large body of work for the hole-doped superconductors to find a profound disagreement between the theoretical expectations and the measurements: this approach predicts a step-like feature rather than a sharp resonance peak, it underestimates the intensity of the resonance by an order of magnitude, it suggests an unreasonable temperature dependence of the resonance, and most severely, it predicts that most of the spectral weight resides in incommensurate wings which are a key feature of the hole-doped cuprates but have never been observed in the electron-doped counterparts. Our findings strongly suggest that the magnetic fluctuations reflect the quantum-mechanical competition between antiferromagnetic and superconducting orders. 
\end{abstract}

\date{\today}

 \maketitle

\section{Introduction}
\label{sec.intro}

High-$T_c$ superconductivity occurs in doped Mott-insulators \cite{Bednorz+86}, electronic states which are insulating because of dominating electron-electron interactions and characterized by spin-only antiferromagnetism.  The study of the fate of this magnetism in the doped, superconducting systems has been on the forefront of high- research from the very beginning with inelastic neutron scattering being the primary source of experimental information.  This is now well documented in the hole doped (p-type) superconductors where one finds the famous "hourglass" spectrum of magnetic fluctuations \cite{Rossat+91,Dai+01,Stock+04,Hayden+04,Christensen+04,Tranquada+04,Woo+06}, but the interpretation of these findings has been subject of severe controversy.  On the one hand this can be interpreted as the signature of strong interaction physics, where the magnetic fluctuations signal the competition between superconductivity and incommensurate, Mott-like antiferromagnetism (the "stripes") \cite{Kivelson+03,Vojta+06}: what matters most in this interpretation is that it is envisaged that the measurements reflect the quantum dynamics of competing, highly collective order parameter fields \cite{Demler+98}. However, it turns out that these data find also a credible interpretation in terms of conventional Fermi-liquid (FL) physics \cite{Norman00,Norman01,Eremin+05,Eschrig06}.  Here, it is asserted that the electron system renormalizes in a weakly interacting Fermi-gas, acquiring a conventional Bardeen-Cooper-Schrieffer (BCS) gap in the superconducting state.  The residual interactions then give rise to weakly bound states residing in the gap formed in the non-interacting particle-hole spectrum, governed by the random phase approximation (RPA).

 Only very recently inelastic neutron scattering data became available for the magnetic fluctuations in the electron doped (n-type) superconductor Pr$_{0.88}$LaCe$_{0.12}$CuO$_4$ (PLCCO) \cite{Wilson+06a}. The spectrum is dominated by a dynamical peak (resonance) at an energy  $\omega_\textrm{res}\approx 11$ meV residing at the antiferromagnetic wavevector $\bqaf=(\pi,\pi)$ whereas the incommensurate branches (wings) found in the p-type superconductors in the vicinity of the resonance are conspicuously absent.  

Here we will employ tunneling spectra obtained for the same sample as used for the neutron measurement, in combination with angular resolved photo emission spectroscopy (ARPES) by others on a similar sample to characterize the fermionic quasiparticle excitations in great detail.  We subsequently use this information to derive the magnetic spectrum employing the RPA, to find out that there is a profound disagreement between the theoretical predictions for the magnetic fluctuations coming from this fermiology interpretation and the measurements \cite{Wilson+06a}.  In particular, (i) this framework predicts a very asymmetric almost step-like feature slightly above the edge of the particle-hole continuum  instead of a sharp resonance peak seen in neutron scattering, (ii) it suggests a strong temperature dependence of the resonance feature, both in intensity and position, inconsistent with the data, (iii) it underestimates the absolute intensity of the resonance by an order of magnitude, and finally (iv) it predicts that most of the spectral weight resides in incommensurate wings below the resonance feature in clear contradiction to the data \cite{Wilson+06a}. 

The outline of this paper is as follows. In Section \ref{sec.rpa} we explain the workings of the FL/RPA approach and the extraction of the quasiparticle parameters from ARPES data and our tunneling experiments. The latter we describe in detail in Section \ref{sec.tunnel}. The results of the theoretical calculations are presented in Section \ref{sec.results} and compared to the magnetic excitation spectrum of PLCCO. Finally, our results and implications of our findings are summarized and discussed in Section \ref{sec.dis}.

\section{Details of the FL/RPA calculations}\label{sec.rpa}

Let us first describe the standard calculations based on the FL/RPA framework. In this approach it is assumed that the cuprates can be interpreted as FLs all along (including the normal state) undergoing a weak coupling BCS instability towards a $d$-wave superconductor, while the excitations are calculated from the leading order in perturbation theory (RPA) controlled by the weakness of the residual interactions. The spin susceptibility within RPA can be written as 

\begin{equation}
\chi(\bq,\omega)=\frac{\chi_0(\bq,\omega)}{1-U(\bq)\chi_0(\bq,\omega)},
\label{eq.rpa}
\end{equation}
where $U(\bq)$ denotes the fermionic four-point vertex and $\chi_0(\bq,\omega)$ the bare non-interacting BCS susceptibility, which is completely determined by the normal state tight-binding dispersion $\epsilon(\bq)$ and the superconducting gap function $\Delta(\bq)$, namely\cite{Bulut+96}

\begin{eqnarray}
\chi_0(\bq,\omega) & = & \sum_\bk\left[\frac 12 (1+\Omega_{\bk,\bq})\frac{f(E_{\bk+\bq})-f(E_\bk)}{\omega-(E_{\bk+\bq}-E_\bk)+i0^+}\right.\nn\\
& & +\frac 14 (1-\Omega_{\bk,\bq})\frac{1-f(E_{\bk+\bq})-f(E_\bk)}{\omega+(E_{\bk+\bq}+E_\bk)+i0^+}\nn\\
& & \left.+\frac 14 (1-\Omega_{\bk,\bq})\frac{f(E_{\bk+\bq})+f(E_\bk)-1}{\omega-(E_{\bk+\bq}+E_\bk)+i0^+}\right].
\label{eq.bcs}
\end{eqnarray}
Here $E(\bq)=\sqrt{\epsilon^2(\bq)+\Delta^2(\bq)}$ denotes the quasiparticle dispersion, $f$ the Fermi function, and for abbreviation we have defined $\Omega_{\bk,\bq}=(\epsilon_{\bk+\bq}\epsilon_\bk+\Delta_{\bk+\bq}\Delta_\bk)/(E_{\bk+\bq}E_\bk)$. The three parts in $\chi_0(\bq,\omega)$ are due to quasiparticle scattering, quasiparticle pair creation and quasiparticle pair annihilation, respectively. 

In the FL/RPA approach for the magnetic resonance mode of the p-type cuprates, the dispersing incommensurate wings merging into the commensurate resonance peak at $\bqaf$ are interpreted as a dispersing bound state formed in the gap below the particle-hole continuum. Such a bound-state corresponds to a pole in the imaginary part of the susceptibility, 
$\chi''(\bq,\omega)$, given by the conditions $1-U(\bq)\chi_0'(\bq,\omega)=0$ and $\chi_0''(\bq,\omega)=0$ for the real and imaginary part of the bare BCS susceptibility $\chi_0(\bq,\omega)$, respectively. The latter condition, the vanishing of the bare Lindhard function $\chi_0''(\bq,\omega)$, enforces a resonance at $(\bq,\omega)$ to live at an energy $\omega$ below the gap of the particle-hole continuum, $\omega<2\Delta(\bq)$.

Before we can employ the RPA formula (\ref{eq.rpa}) to calculate the magnetic response $\chi(\bq,\omega)$, we have to characterize the bare quasiparticles in great detail to determine the bare BCS susceptibility (\ref{eq.bcs}). In particular, we have to use experimental input to extract the normal state dispersion $\epsilon(\bq)$ and the superconducting gap $\Delta(\bq)$.  For the normal state dispersion we use the standard square lattice tight-binding dispersion

\begin{eqnarray}
\epsilon(\bq) & = & -2t[\cos(k_x)+\cos(k_y)]-4t'\cos(k_x)\cos(k_y)\nn\\
& & -2t''[\cos(2 k_x)+\cos(2 k_y)]\nn\\
& & -4t'''[\cos(2 k_x)\cos(k_y)+\cos(k_x)\cos(2 k_y)]\nn\\
& &-4t^{iv}\cos(2 k_x)\cos(2 k_y)-\mu,
\end{eqnarray}
having incorporated an appropriate chemical potential $\mu$. A normal state single particle dispersion $\epsilon(\bq)$
for optimally doped PLCCO of this form has been determined\cite{Das+06} by fitting the ARPES data at 30 K \cite{Matsui+05} along three independent directions. The resulting tight-binding parameters are listed in Table \ref{par}.

\begin{table}[htbp]
  \begin{center}
    \begin{tabular}{c||c|c|c|c|c|c|c|c|c|c}
       & $t$ & $t'$ & $t''$ & $t'''$ & $t^{iv}$ & $\mu$ & $\Delta_1$ & $\Delta_3$ & $U$ & $\Delta U$  
      \\\hline
      YBCO & 250 & -100 & 0 & 0 & 0 & -270.75 & 42 & 0 & 572 & 57.2  
      \\\hline
      PLCCO & 120 & -60 & 34 & 7 & 20 & -82 & 5.44 & 2.24 & 500 & 0 
      \end{tabular}
   \end{center}
\caption{Collection of parameters used in our calculation: parameters of the normal state tight-binding dispersion $t,t',t'',t''',t^{iv}$, chemical potential $\mu$, $d$-wave gap parameters $\Delta_1,\Delta_3$, and four point vertex parameters $U,\Delta U$. Tight-binding parameters for PLCCO are taken from Ref. \onlinecite{Das+06} and parameters for YBCO from Ref. \onlinecite{Eremin+05}.} 
\label{par}
\end{table}

\begin{figure}[ht]
\epsfig{figure=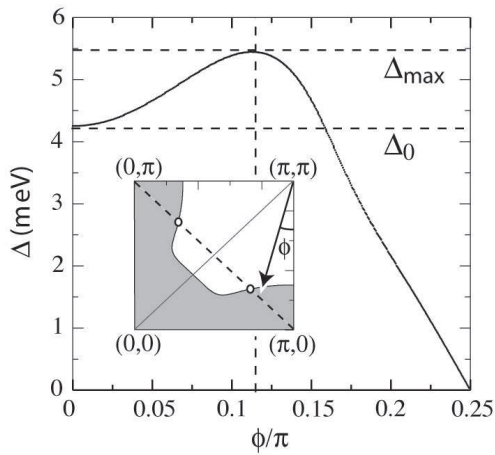,width=0.75\linewidth}
\caption{The non-monotonic $d$-wave gap $\Delta$ of PLCCO along the Fermi surface as a function of the Fermi surface angle $\phi$ (see inset) calculated with the set of parameters listed in Table \ref{par}. The inset shows the relation between the Fermi surface and the magnetic Brillouin zone. The hot spots relevant for the magnetic response at $\bqaf$ are shown as open circles. The position of the gap maximum close to the hot spots and the ratio $\Delta_\textrm{max}/\Delta_0$ of the maximum gap value and the antinodal gap are in good agreement with ARPES measurements\cite{Matsui+05}. The absolute gap values are extracted from our tunneling experiment (see Section \ref{sec.tunnel}).}
\label{gap}
\end{figure}

To reproduce the non-monotonic $d$-wave gap of PLCCO observed in the ARPES measurement\cite{Matsui+05}, we include third harmonics in the gap function, 

\begin{eqnarray}
\Delta(\bk) & = & \frac{\Delta_1}{2}[\cos(k_x)-\cos(k_y)]\nn\\
& & -\frac{\Delta_3}{2}[\cos(3 k_x)-\cos(3 k_y)]
\end{eqnarray}
and adjust the ratio $\Delta_1/\Delta_3$ to reproduce the functional form of the gap along the Fermi surface found experimentally. For $\Delta_1/\Delta_3\approx 2.43$ we find a maximum gap value $\Delta_\textrm{max}\approx 1.3\Delta_0$ under a Fermi surface angle $\phi_\textrm{max}\approx 21^\circ$ with  $\Delta_0$ the gap value at the anti-nodal direction ($\phi=0$) in agreement with the experimental observation (see Fig. \ref{gap}). The gap maxima are very close to the intersection points of the Fermi surface and the magnetic Brillouin zone $|k_x|+|k_y|\leq\pi$. These so called hot spots are relevant for particle-hole processes contributing to the magnetic response at  $\bqaf$.  

To determine absolute gap values which are difficult to extract from leading-edge shifts in ARPES data we have performed tunneling experiments on the same sample of PLCCO showing a magnetic resonance at $\omega_\textrm{res}\approx 11$ meV\cite{Wilson+06a}. A detailed discussion of the experimental setup and the obtained results are presented in section \ref{sec.tunnel}. 

The complete set of parameters for the normal-state dispersion $\epsilon(\bq)$ and the gap function $\Delta(\bq)$ of PLCCO is listed in Table \ref{par} and compared to a set of parameters used recently to calculate the magnetic response of optimally doped YBCO\cite{Eremin+05}. The latter we use as a benchmark for our numerical calculation and also for a comparison of the features of the FL/RPA spectra of the n- and p-type compounds.   

To calculate the bare susceptibility $\chi_0(\bq,\omega)$, we replace $i0^+$ by $i\Gamma$  in the energy denominators, mimicking experimental broadening. We take $\Gamma=2$ meV consistent with the typical broadening in neutron scattering and values used in other RPA calculations. The resulting well-behaved function is then summed numerically over a 1500 by 1500 mesh in the Brillouin zone.

Since the bare non-interacting BCS susceptibility $\chi_0(\bq,\omega)$ is completely determined by $\epsilon(\bq)$ and $\Delta(\bq)$, we can only adjust $U(\bq)$ in the RPA equation (\ref{eq.rpa}) to reproduce the magnetic excitation spectrum of PLCCO. Following other standard RPA calculations for p-type compounds, we take an onsite repulsion $U_0$ and allow for a small $\bq$-modulation with amplitude $\Delta U$ (see e.g. Ref. \onlinecite{Eremin+05}), $U(\bq)=U_0-\Delta U [\cos(q_x)+\cos(q_y)]$.

\section{Tunneling experiment}\label{sec.tunnel}

\begin{figure}[ht]
\epsfig{figure=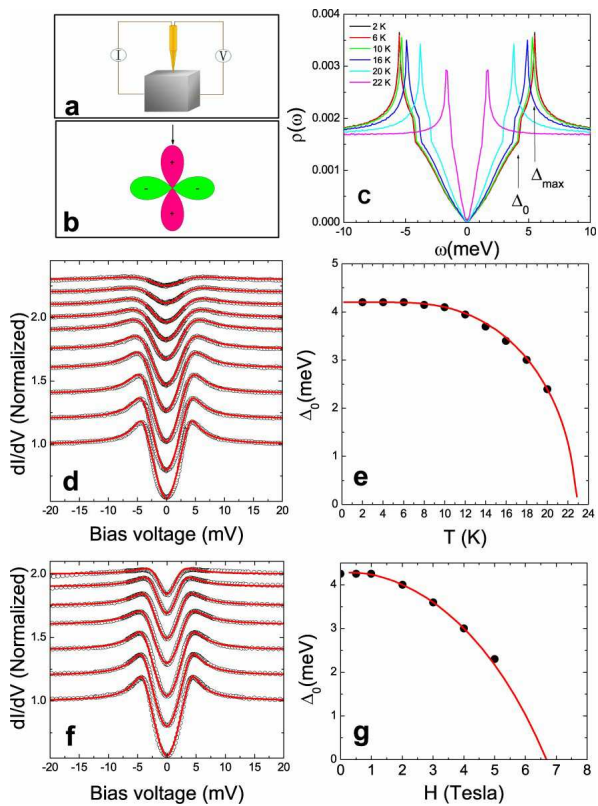,width=0.9\linewidth}
\caption{Geometry and results of direct point-contact tunneling measurements on single crystals of PLCCO. (a) The schematic diagram of the experimental setup, where a Au tip is pointed towards the $a/b$ axis direction determined by neutron diffraction. (b) The relationship between standard $d$-wave gap and tunneling direction. (c) Calculated quasiparticle density of states using gap values at different temperatures showing Van Hove singularities at the antinodal gap $\Delta_0$ and the maximum gap $\Delta_\textrm{max}$. (d) Temperature dependence of the $dI/dV$ spectra from 2 K to 20 K every 2 K. The spectra were obtained by normalizing the corresponding backgrounds at temperatures well above $T_c$. (e) Temperature dependence of the gap value $\Delta_0$, the solid line denotes the BCS prediction. (f) Magnetic field dependence of the $dI/dV$ spectra for a $c$-axis aligned magnetic field. The theoretical calculations are indicated by red lines in (d) and (f), respectively. All the spectra and fitting lines except for the lowest ones are shifted upwards for clarity. (g) Superconducting gap as a function of increasing magnetic field, the solid line is the guide to eyes. The $\Delta_0$ values in (e) and (g) are determined by fitting the normalized spectra to the extended Blonder-Tinkham-Klapwijk model\cite{Blonder+82} with a $d$-wave-type gap function along the $a/b$ axes.}
\label{tunnel}
\end{figure}

To determine the absolute gap value of PLCCO and its temperature dependence, we performed tunneling measurements on the same sample used for the neutron scattering measurements\cite{Wilson+06a}. The directional point-contact tunneling measurements were carried out by pointing a Au tip towards the specified directions of $a$ or $b$ crystal axis which is determined by neutron scattering (Fig. \ref{tunnel} a,b). The Au tips were mechanically sharpened by carefully clipping a gold wire with a diameter of 0.25mm. The approaching of the tips were controlled by a refined differential screw. The point contact insert was set in the sample chamber of an Oxford cryogenic system Maglab-EXA-12. In order to reduce the quasiparticle scattering in the barrier layer and hence obtain high quality data, the nonaqueous chemical etch was used to attenuate the insulating layer on the sample surface immediately before mounting the sample on the point contact device\cite{Shan+05}. 

Typical four-terminal and lock-in techniques were used to measure the $I\sim V$ curves and the differential resistance $dV/dI$ vs $V$ of the point contacts simultaneously. Then the dynamical conductance $dI/dV\sim V$ was obtained both by converting the $dV/dI\sim V$ curves and by calculating the derivative of $I\sim V$ relations in order to ensure the reliability of the results. It was verified that the results were not affected by the heat-relaxation effect by comparing the curves recorded by positively and negatively bias scanning. For quantitative analysis, the spectra were normalized by corresponding backgrounds constructed according to the spectrum measured well above $T_c$. 

In Fig. \ref{tunnel}d, we show the temperature dependence of the $dI/dV$ spectra from 2 K to 20 K with increments of 2 K. Note that due to the experimental broadening the two van-Hove singularities at $\Delta_0$ and $\Delta_\textrm{max}$ (see Fig.\ref{gap}) in the density of states $\rho(\omega)$ (Fig.2c) are not resolved. To make it as advantageous as possible for the the FL/RPA approach to explain the magnetic resonance, we identify the gap seen in the tunneling spectra with the gap $\Delta_0$ at the antinodal direction. 
This probably overestimates the true gap since from the data we probably extract an energy between $\Delta_0$ and $\Delta_\textrm{max}$. On the other hand, we note that  
point-contact tunneling measures the density-of-states averaged superconducting gap, its value might be different from those obtained by spatially resolved scanning tunneling microscopy.

From a fit to the extended Blonder-Tinkham-Klapwijk (BTK) model \cite{Blonder+82} with a $d$-wave-type gap function \cite{Shan+05}, we obtain the BCS like temperature dependence of the gap value as shown in Fig. \ref{tunnel}e. Similarly, from the dependence of the  spectra on $c$-axis aligned magnetic field, we extract the superconducting gap as a function of increasing magnetic field (Fig. \ref{tunnel}f,g).

\section{RPA results and comparison to experiments}\label{sec.results}

\begin{figure}[ht]
\epsfig{figure=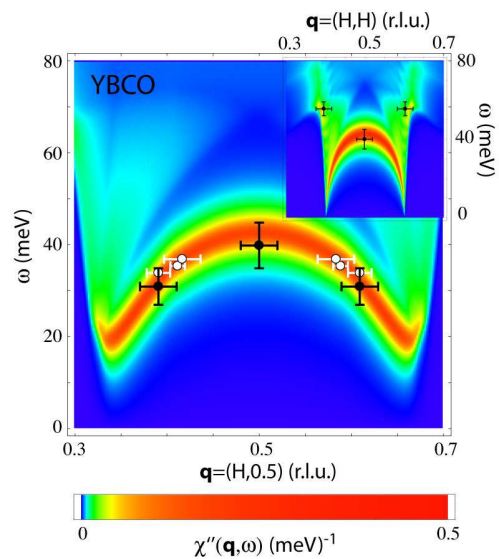,width=0.75\linewidth}
\caption{Magnetic response $\chi''(\bq,\omega)$ calculated within the FL/RPA approach using a set of parameters optimized for optimally doped YBCO\cite{Eremin+05} plotted along the $(H,0.5)$ and $(H,H)$ (inset) directions close to $\bqaf$.  White and black points show neutron scattering data from Ref. \onlinecite{Dai+01} and Ref. \onlinecite{Woo+06}, respectively.}
\label{YBCO}
\end{figure}

Before we calculate the magnetic response $\chi''(\bq,\omega)$ for PLCCO within the FL/RPA framework using the tight-binding dispersion $\epsilon(\bq)$ and the gap function $\Delta(\bq)$ determined by ARPES\cite{Matsui+05} and our tunneling experiment, we first test our numerical routine for a set of parameters that has been used to calculate the magnetic excitation spectrum of optimally doped YBCO\cite{Eremin+05}. The resulting magnetic excitation spectrum in the vicinity of the antiferromagnetic wave vector is shown in Fig. \ref{YBCO} along the $(H,1/2)$ and $(H,H)$ directions, respectively and is found to be in perfect agreement with the theoretical results in Ref. \onlinecite{Eremin+05}. The favorable comparison of theoretical results with the dispersion found in inelastic neutron scattering experiments\cite{Dai+01,Woo+06} on optimally doped YBCO is also shown in Fig. \ref{YBCO}. However, a closer inspection of the intensities shows that the FL/RPA calculation severely underestimates the spectral weight above the commensurate dynamical resonance\cite{Reznik+06}. Whereas experimentally the intensities of the upper and lower wings forming the characteristic hourglass in the vicinity of the resonance are quite comparable\cite{Woo+06} in the RPA results the upper half of the hourglass is completely absent (see Fig. \ref{YBCO}).

\subsection{Resonance feature of PLCCO}\label{sec.res}

Before we calculate the full momentum dependent RPA spectrum of PLCCO, we try to reproduce the resonance feature at $\omega_\textrm{res}\approx 11$meV found in inelastic neutron scattering\cite{Wilson+06a} by tuning the value of the four-point vertex $U=U(\bqaf)$ at the antiferromagnetic wavevector.  

\begin{figure}[h]
\epsfig{figure=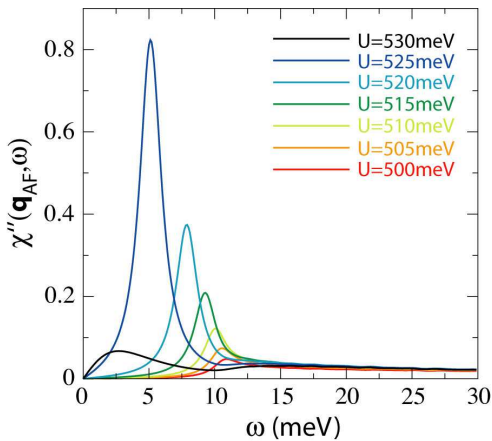,width=0.75\linewidth}
\caption{Evaluation of  $\chi''(\bqaf,\omega)$ in the superconducting state for different values of $U$ showing the narrow energy window $515$meV$<U<528$meV for which a bound state in the gap $2\Delta=10$meV of the particle-hole is formed. For $U=500$meV we find an intensity enhancement at the experimentally observed resonance energy $\omega_\textrm{res}=11$meV. Since the feature is located slightly above the gap the intensity is significantly reduced compared to the bound-state situation and the line-shape is very asymmetric. }
\label{Jdep}
\end{figure}

The smallness of the gap $\Delta\approx 5$meV relevant for $\bqaf$-scattering (see Fig. \ref{gap}) enforces a worrisome fine-tuning to produce a bound state. The necessary conditions for the corresponding singularity in the imaginary part  of the dynamic susceptibility $\chi''(\bqaf,\omega)$ are given by $\omega<2\Delta$ and $U=1/\chi_0'(\bqaf,\omega)$. The evolution of $\chi''(\omega)$ for different values of $U$ is plotted in Fig. \ref{Jdep}. For $U<515$meV the resonance is pushed into the particle-hole continuum whereas the system runs into a magnetic instability for $U>528$meV.  Since experimentally the resonance peak is found slightly above the edge of the particle-hole continuum as confirmed by our tunneling measurements (Fig. \ref{tunnel}e,g) it cannot be explained as a bound state.  For $U=500$meV, we find an intensity enhancement around 11meV. As expected, since no bound state is formed the FL/RPA result has a very asymmetric and almost step-like line-shape, rather than the symmetric peak observed in experiment  \cite{Wilson+06a} (see Fig. \ref{absint}b), and the intensity is significantly reduced compared to a typical bound state situation. 

\subsection{Temperature dependence}\label{sec.tdep}

In this section we are going to analyze the temperature dependence of the resonance feature that is to be expected in the FL/RPA framework taking the BCS like temperature dependence of the gap given by the tunneling experiment. Since the two features in the quasiparticle density of states at the antinodal gap $\Delta_0$ and the maximum gap $\Delta_\textrm{max}$ are not resolved in the data (see Fig. \ref{tunnel}c,d) we assume the non-monotonic functional form of the gap along the Fermi surface not to change with temperature and simply scale the gap function plotted in Fig. \ref{gap} according to BCS like temperature dependence extracted from the tunneling data (Fig. \ref{tunnel}e). 

\begin{figure}[h]
\epsfig{figure=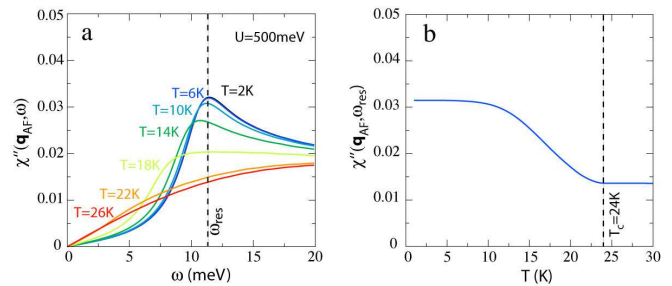,width=\linewidth}
\caption{(a)  Evolution of the resonance feature calculated for PLCCO within the FL/RPA approach with temperature. With increasing temperature the resonance feature shifts continuously to lower energy and decreases in intensity. (b) Calculated intensity at $\omega_\textrm{res}$ as a function of temperature showing a strong decrease at temperatures well below $T_c$.}
\label{Tdep}
\end{figure}

The resulting temperature dependence of the "resonance" feature we obtained at zero temperature at 11meV for $U=500$meV  is summarized in Fig. \ref{Tdep}. From the calculation, below $T_c\approx24$K we expect a strong temperature dependence of the resonance feature both in position and intensity.  With increasing temperature the resonance shifts to lower energies whereas the intensity goes down continuously (Fig. \ref{Tdep}a,b). These predictions are inconsistent with the experimental observations, where the position of the resonance appears to be fixed and the intensity drops down sharply close to $T_c$\cite{Wilson+06a}. 

Since the temperature dependence of the gap is expected to be the dominant effect we have not taken thermal broadening into account. However, additional broadening would even lead to a stronger continuous decrease of the intensity below $T_c$.

\subsection{Comparison of absolute intensities}\label{sec.int}

Since the magnetic resonance of PLCCO is too high in energy it cannot be explained as a bound state within the FL/RPA framework and has consequently to be identified with a weak intensity enhancement slightly above the gap of the particle-hole continuum. Therefore we expect a significant reduction of the spectral weight compared to a typical bound-state situation. Comparing the intensity of the "resonance" feature we produced within the RPA calculation for $U=500$meV at $\omega_\textrm{res}=11$meV with the intensity of the bound state one obtains with the set of parameters optimized for YBCO\cite{Eremin+05} (see Table \ref{par}), we expect the resonance of PLCCO to be weaker by a factor 15 than the resonance of YBCO. In this comparison we have used the same broadening $\Gamma=2$meV for both cases.   

To compare this theoretical expectation with experiment, we have converted the neutron scattering raw data on the resonance of PLCCO reported in Ref. \onlinecite{Wilson+06a}  to absolute units ($\mu_B^2$eV$^{-1}$f.u.$^{-1}$) both in the normal and superconducting states by normalizing them to acoustic phonons around the (2,0,0) Bragg reflection \cite{Stock+04}. 

In the long-wavelength limit, the differential cross section for coherent one phonon emission at given $(\vec{\kappa},\omega)$ is\cite{Shirane+02} 

\begin{eqnarray}
\frac{\partial^2\sigma}{\partial\Omega\partial E} & = & A\frac{\hbar^2 N}{2 E(\bq)}\frac{k_f}{k_i}(n(\omega)+1)(\vec{\kappa}\cdot\hat{e}_{\bq s})^2\nn\\
& & \times e^{-2W}\frac 1M |G(\vec{\tau})|^2\delta(E-E(\bq)),
\end{eqnarray}
where $\vec{\kappa}=\vec{\tau}+\bq$ is the momentum transfer of the neutron, $E(\bq)$ the energy of the phonon mode, $N$ the number of unit cells, $k_i$ and $k_f$ are the incident and final wavelengths of the neutron, $n(\omega)$ is the standard Bose population factor, $\hat{e}_{\bq s}$ is the unit vector in the direction of atomic displacement for the phonon mode, $e^{-2W}$ is the Debye-Waller factor, $M$ the mass of the unit cell, and $G(\vec{\tau})$ is the standard nuclear structure factor. The spectrometer dependent constant $A$ can be determined through the measurement of a known phonon in the material. For our case, we measured a transverse acoustic phonon at $\bQ=(0.12, 2, 0)$.

The same spectrometer dependent constant $A$ can then be used to determine the magnetic susceptibility in absolute units. For paramagnetic spin fluctuations the cross section is 

\begin{eqnarray}
\frac{\partial^2\sigma}{\partial\Omega\partial E} & = & A\frac{(\gamma r_0)^2}{4}\frac{k_f}{k_i}N|f(\vec{\kappa})|^2(n(\omega)+1)\nn\\
& & \times e^{-2W}\frac{2}{\pi\mu_B^2}\chi''(\vec{\kappa},\omega),
\end{eqnarray}
where $(\gamma r_0)^2/4$ is $7.265\cdot 10^{-26}$ cm$^2$ and $f(\vec{\kappa})$ is the isotropic, magnetic form factor for Cu$^{2+}$. In order to obtain the local susceptibility $\chi''(\omega)=V_Q^{-1}\int\chi''(\bQ,\omega)\ud^3Q$ at the $(\pi,\pi)$ in-plane wavevector, $Q$-scans were performed at selected energies. For energies below $5$ meV and above $10$ meV, SPINS data and BT-9 data were respectively cross-normalized to the absolute values of the HB-1 data using constant scale factors. For energies below $5$ meV, the measured $Q$-widths along $[H,H]$ were broader than resolution while scans at all higher energy transfers all showed resolution limited peaks along $[H,H]$. In order to estimate the local susceptibility, the magnetic signal was assumed to be a two-dimensional Gaussian within the $[H,K]$ plane and rod-like out of plane. This neglects the rotation of the resolution ellipsoid at energy transfers away from the resonance position and results in a slight underestimation of the integrated magnetic scattering at energies below the resonance. This estimation however is systematic and does not influence relative changes in the local susceptibility as the system enters the superconducting phase. For points in $E$-scans with $E>5$ meV in which no $Q$-scan data was available, the calculated resolution value was used projected along the $[H,H]$ direction. The background was removed through subtracting the measured nonmagnetic signal away from the correlated $(\pi,\pi)$ position as shown in Ref. \onlinecite{Wilson+06a}. All data was corrected for $\lambda/2$ contamination in the monitor, and in our calculations for data at both $2$ K and $30$ K, the Debye-Waller factor was assumed to be 1.

\begin{figure}[h]
\epsfig{figure=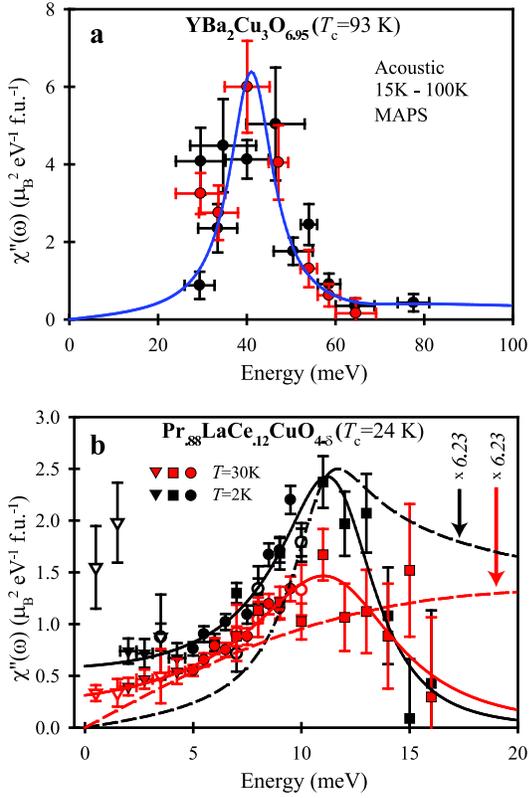,width=0.8\linewidth}
\caption{Comparison of the resonance in absolute units with FL/RPA calculations for optimally hole-doped YBCO and electron-doped PLCCO. (a) Local susceptibility in absolute units for optimally doped YBCO at 10 K from Ref. \onlinecite{Woo+06}. The solid blue line is the calculation based on RPA model scaled to match the experimental data.  (b) Local susceptibility in both the normal and superconducting state for PLCCO obtained from converting the raw data of Ref. \onlinecite{Wilson+06a} to absolute units.  Solid lines are guide to the eyes. The dashed lines represent the results of the FL/RPA calculations with the same scale factor as used for YBCO. Note that the theoretical values are about 6 times smaller than the experimental results.}
\label{absint}
\end{figure}

Assuming all the scattering centered at ${\bf Q}=(1/2,1/2,0)$ is magnetic, 
we find that the local susceptibility $\chi''(\omega)$ has a peak around 11 meV and increases at all energies probed (from 0.5 meV to 16 meV) on cooling from the normal state to the superconducting state (Fig. \ref{absint}b). This is in contradiction to the theoretical results which predict a reshuffling of spectral from low energies in the normal state to the resonance feature in the superconducting state (Figs. \ref{Tdep}a,\ref{absint}b). 

Figure \ref{absint}a shows the local susceptibility for optimally doped YBCO \cite{Woo+06}, where the resonance intensity is obtained by taking the temperature difference between the normal (100 K) and superconducting states (10 K) since the absolute intensity of the mode in the normal state is still unknown. 
In Figure \ref{absint}b we plot the local susceptibility in the normal (30 K) and superconducting (2 K) states
normalized to phonons. The local susceptibility in absolute unit is similar to those of PLCCO with a different $T_c$\cite{Wilson+06b} and Pr$_{0.89}$LaCe$_{0.11}$CuO$_4$\cite{Fujita+06}, and
is about 2.5 times smaller than that of the resonance for YBCO in Figure \ref{absint}a.
  From the FL/RPA calculations we expect the spectral weight of the resonance to be smaller by a factor 15 as compared to YBCO's resonance (see Fig. \ref{absint}), in clear contrast to experiments. 

\subsection{Momentum dependence}\label{sec.qdep}

Finally, we calculate the momentum dependence of the imaginary part $\chi''(\bq,\omega)$ of the dynamic susceptibility in the vicinity of $\bqaf$ using the band structure parameters and superconducting gap discussed in Sec. \ref{sec.rpa} as appropriate for PLCCO. In Sec. \ref{sec.res} we have seen that for $U(\bqaf)=500$ meV the FL/RPA approach reproduces a feature at $\omega_\textrm{res}=11$ meV. However, since this feature is located at an energy above the gap of the particle-hole continuum its lineshape and spectral weight are inconsistent with the experimental observations.   

We start with a momentum independent four-point vertex $U(\bq)=U$ (Hubbard-like approximation) which in the the case of p-type compounds turns out to give a pretty good description of the magnetic excitation spectra\cite{Norman00,Norman01}, much better than for a strong momentum dependent $U(\bq)$\cite{Norman00}.  

\begin{figure}[h]
\epsfig{figure=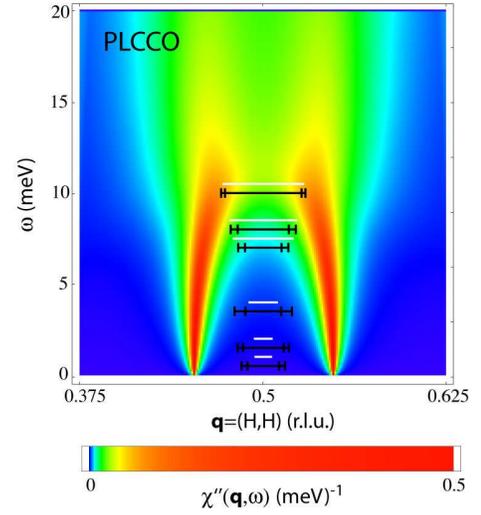,width=0.7\linewidth}
\caption{Comparison of the magnetic excitation spectrum $\chi''(\bq,\omega)$ along the $[H,H]$ direction in the vicinity of $\bqaf=(1/2,1/2)$r.l.u. resulting from the FL/RPA calculations with neutron scattering data\cite{Wilson+06a} on optimally doped PLCCO ($T_c=24$K) measured at $T=2$K. The very strong incommensurate wings predicted by the calculations highlight the failure of the FL/RPA approach.}
\label{PLCCO}
\end{figure}

Using a constant $U=500$ meV producing a resonance feature at the experimentally observed energy $\omega_\textrm{res}=11$ meV\cite{Wilson+06a} the FL/RPA predicts for the n-type superconductor the spectrum shown in Figure \ref{PLCCO}: this spectrum is dominated by strong incommensurate wings below the resonance which are in fact predicted to be much more pronounced than in the case of the p-type superconductors. This is precisely opposite to the experimental findings where the incommensurate fluctuations are pronounced in the p-type systems, but completely absent in the n-type superconductor!  

Including a small $\bq$-modulation of the form $U(\bq)=U_0-\Delta U[\cos(q_x)+\cos(q_y)]$ as used recently\cite{Eremin+05} with a relative modulation $\Delta U/U_0=0.1$ to obtain a slightly better quantitative agreement with continuously improving neutron scattering data on optimally doped YBCO, does not lead to significant improvements but only to a small change of the incommensurability of the wings.

The only way to repair this gross inconsistency is by invoking a $U(\bq)$ which sharply peaks at $\bqaf$. Recently, it was argued\cite{Ismer+07} that by taking a full momentum dependent four-point vertex ($U_0\to0$), $U(\bq)=-J [\cos(q_x)+\cos(q_y)]/2$, the incommensurate wings can be suppressed. This strongly momentum-dependent form of the four-point vertex peaking at $\bqaf$ was motivated by the proximity of the superconducting and (commensurate) antiferromagnetic phases. However, such a form of the four-point vertex is clearly unphysical since it corresponds to a nearest neighbor exchange whereas the onsite Coulomb repulsion which is known to control the Mottness in the copper oxide planes is completely ignored. In contrast to the $tJ$-model, in the quasiparticle picture used here double occupancies are not projected out. Moreover, using our set of quasiparticle parameters (Tab. \ref{par}), this would imply an effective superexchange of $J=500$ meV which is about 5 times bigger than in the parent undoped compounds\cite{Bourges+97}. While this is obviously unphysical, the value $J=854$ meV taken in Ref. \onlinecite{Ismer+07} is even much bigger.

\begin{figure}[h]
\epsfig{figure=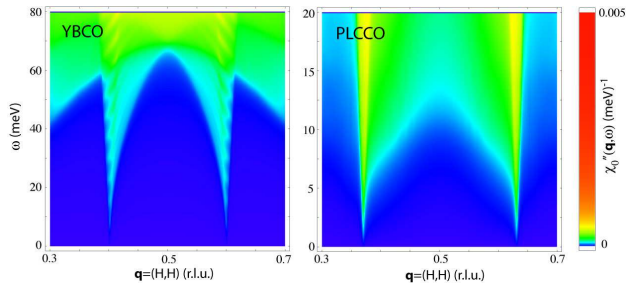,width=0.95\linewidth}
\caption{Bare Lindhard functions $\chi_0''(\bq,\omega)$ of YBCO and PLCCO in the superconducting phases calculated with bandstructure and $d$-wave gap parameters listed in Table \ref{par}. Whereas the momentum dependence of the gap of the ph-continuum looks very similar for the p-type and n-type material, the distribution of spectral weight is completely different. For the p-type, spectral weight is accumulated at $\bqaf$ whereas for the n-type a lot of intensity has shifted from $\bqaf$ to incommensurate wavevectors.}
\label{chi0}
\end{figure}

\begin{figure}[h]
\epsfig{figure=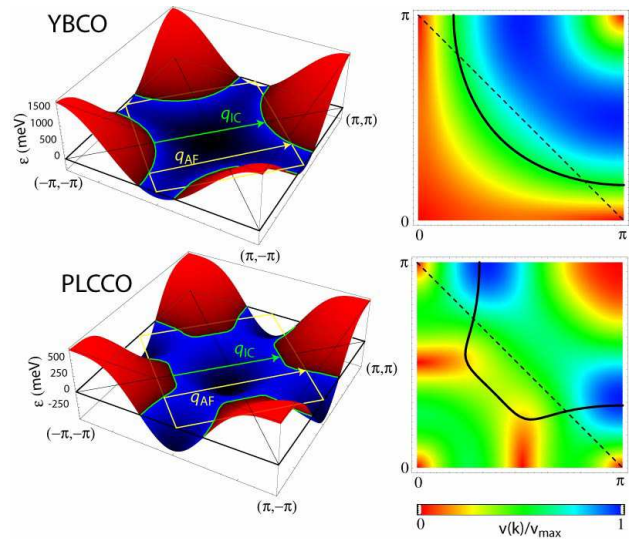,width=0.95\linewidth}
\caption{Comparison of the normal state dispersions, Fermi surfaces (left panel) and Fermi velocities (right panel) of YBCO and PLCCO.}
\label{disp}
\end{figure}

The reason why incommensurate wings at low energies appear generically within the FL/RPA approach for any realistic set of parameters both for p-type and n-type materials is actually a very generic one, rooted in the assumption that there is a direct relation between the free particle-hole and the magnetic spectrum. Within this framework, the RPA response $\chi''(\bq,\omega)$ for any realistic form of $U(\bq)$ basically reflects the momentum dependence of the gap of the ph-continuum nicely seen in the bare Lindhard function $\chi_0''(\bq,\omega)$ (Fig. \ref{chi0}). The superconducting $d$-wave gap is close to its maximum for particle-hole pairs separated by $\bqaf$ and goes continuously down if we move away from the antiferromagnetic to incommensurate wavevectors separations (see Figs. 
\ref{gap},\ref{disp}). The gap of the ph-contiuum closes at the incommensurate wavevectors connecting points of the Fermi surface coinciding with the nodes of the $d$-wave gap.   

Although the momentum dependence of the gap of the particle-hole continuum looks very similar in the p-type and n-type case, a crucial difference becomes apparent when comparing the distribution of the spectral weight $\chi_0''(\bq,\omega)$. Whereas in p-type YBCO spectral weight is accumulated at $\bqaf$ the intensity in the close vicinity of the antiferromagnetic wavevector is strongly suppressed in n-type PLCCO. On the other hand the spectral weight at incommensurate momenta is strongly enhanced in the n-type compound (Fig. \ref{chi0}).

The reason for the reshuffling of the weight in the bare Lindhard function $\chi_0''(\bq,\omega)$ from $\bqaf$ to incommensurate wave vectors in going from p- to n-type superconductors is simply related to the number of particle-hole pairs contributing to the magnetic response. In Fig. \ref{disp} we compare the normal state dispersions, Fermi surfaces, and Fermi velocities of YBCO and PLCCO.  
Whereas in YBCO the saddle-points in the band structure responsible for the van Hove singularities at the antinodal points are very close to points on the Fermi surface separated by $\bqaf$, in PLCCO the bands are very steep at points connected by $\bqaf$ and a nesting of the Fermi surface for incommensurate wavevectors in regions of very flat bands give rise to the drastic spectral weight enhancement of the wings. 

\section{Discussion and conclusion}\label{sec.dis}

To summarize, by combining experimental measurements of the quasiparticle and dynamical magnetic properties we have demonstrated that in a n-type cuprate superconductor the magnetic excitations to be expected from a weakly interacting Fermi-gas are inconsistent with experimental observations. In particular, we have performed tunneling experiments on the same sample of PLCCO showing a magnetic resonance in the superconducting phase \cite{Wilson+06a} and used ARPES data\cite{Matsui+05} on a similar sample to extract normal-state band structure and $d$-wave gap parameters.  The neutron scattering raw data on the magnetic resonance\cite{Wilson+06a} we have converted to absolute units by normalization to acoustic phonons.

Using the detailed information on the quasiparticles obtained from the ARPES and tunneling experiments we have calculated the expected magnetic excitation spectrum within the conventional FL/RPA framework which assumes that there is a direct relation between the free particle-hole and the magnetic spectrum. The comparison of the theoretical results with the magnetic fluctuation measured in inelastic neutron scattering shows that the fermiology approach fails to explain the magnetic fluctuations.

Since the magnetic resonance of PLCCO is located at an energy near the gap of the particle-hole continuum as confirmed by our tunneling experiment, it is difficult to explain it as a bound state within the FL/RPA approach. Consequently, within the FL/RPA framework we obtain an almost step-like feature rather than a symmetric resonance peak seen in experiment and underestimate the spectral weight of the resonance by an order of magnitude.
Additionally, taking the temperature dependence of the gap measured by our tunneling experiment, the FL/RPA approach predicts a very strong temperature dependence of the resonance well below $T_c$ inconsistent with the experimental observation. The failure of the fermiology framework is highlighted by the incommensurate wings which from the theoretical calculations are expected to be much more pronounced than in the p-type case whereas they have never been observed in electron-doped superconductors.

Within the FL/RPA approach such incommensurate wings in the magnetic response of a $d$-wave superconductor appear generically for any physically reasonable set of parameter, both in the p-type and n-type case. This finding is quite robust and does not depend on details of the bandstructure. However, the different forms of the quasiparticle dispersion and of the $d$-wave gap of PLCCO compared to YBCO gives rise to an additional reshuffling of spectral weight in the free particle-hole spectrum from the antiferromagnetic to incommensurate wavevectors leading to an enhancement of the wings and an additional intensity loss in the close vicinity of $\bqaf$.

The drastic failure of the fermiology approach for the n-type case opens the question whether the apparent agreement for the p-type superconductors is just coincidental. Since incommensureate wings are generically expected within the FL/RPA approach it is not surprising that one finds a reasonable agreement up to the resonance energy. However, this approach cannot explain the upper branches of the hourglass spectrum seen in various experiments. 
More severely, the fermiology interpretation can neither account for the anomalous properties of the normal state which is known to be a non-Fermi liquid nor for the persistence of the resonance and the hourglass above $T_c$ in the underdoped regime.   
 
On the other hand, above the spin gap the magnetic excitation spectra of superconducting YBCO\cite{Stock+04,Hayden+04} and La$_{2-x}$Sr$_x$CuO$_4$ \cite{Christensen+04} are remarkably similar\cite{Hayden+04,Tranquada+06} to that found in stripe ordered La$_{1.875}$Ba$_{0.125}$CuO$_4$ \cite{Tranquada+04} suggesting that the magnetic fluctuations in the p-type superconductors correspond to fluctuating stripes competing with superconductivity. Theoretically, the hourglass spectrum characteristic for both, stripe ordered and superconducting p-type cuprates, has been obtained in various models for static stripes\cite{Kruger+03} but also in a phenomenological lattice model for thermally fluctuating, short-ranged stripe order\cite{Vojta+06}.   

Whereas the magnetic fluctuations in the p-type cuprates seem to reflect the competition between superconductivity and incommensurate, Mott-like antiferromagnetism (the "stripes"), incommensurate fluctuations have never been observed in n-type superconductors suggesting instead a competition with commensurate antiferromagnetism\cite{Demler+98}.
 
To conclude, by combining experimental measurements of the quasiparticle and dynamical magnetic properties we have demonstrated that in the n-type cuprate superconductor PLCCO there is no relation whatsoever between  the magnetic excitations to be expected from a weakly interacting Fermi-gas and the magnetic fluctuations observed experimentally. 
This demonstrates that the magnetic fluctuations actually correspond with highly collective motions which likely reflect the quantum competition between superconductivity and strongly coupled antiferromagnetism.  The challenge for the theorist is to explain how this system manages to simultaneously support conventional looking fermionic quasiparticle excitations and highly collective order parameter fluctuations.

\begin{acknowledgments}
The authors like to thank Dirk Morr and Ilya Eremin for stimulating discussions. This work is supported in part by the US National Science Foundation with Grant No. DMR-0453804 and DMR-0342832, by Dutch Science Foundation NOW/FOM, and the US DOE BES under the contract No DE-AC03-76SF00515.  The PLCCO single crystal growth at UT is supported by the US DOE BES under contract No. DE-FG02-05ER46202.  ORNL is supported by the US DOE Grant No. DE-AC05-00OR22725 through UT/Battelle LLC.  The work at IOP, CAS is supported by NSFC, the MOST of China (973 project: 2006CB601000, 2006CB0L1002) and CAS project: ITSNEM.
\end{acknowledgments}


\begin{thebibliography}{43}
\expandafter\ifx\csname natexlab\endcsname\relax\def\natexlab#1{#1}\fi
\expandafter\ifx\csname bibnamefont\endcsname\relax
  \def\bibnamefont#1{#1}\fi
\expandafter\ifx\csname bibfnamefont\endcsname\relax
  \def\bibfnamefont#1{#1}\fi
\expandafter\ifx\csname citenamefont\endcsname\relax
  \def\citenamefont#1{#1}\fi
\expandafter\ifx\csname url\endcsname\relax
  \def\url#1{\texttt{#1}}\fi
\expandafter\ifx\csname urlprefix\endcsname\relax\def\urlprefix{URL }\fi
\providecommand{\bibinfo}[2]{#2}
\providecommand{\eprint}[2][]{\url{#2}}

\bibitem[{\citenamefont{Bednorz and M\"uller}(1986)}]{Bednorz+86}
\bibinfo{author}{\bibfnamefont{J.~G.} \bibnamefont{Bednorz}} \bibnamefont{and}
  \bibinfo{author}{\bibfnamefont{K.~A.} \bibnamefont{M\"uller}},
  \bibinfo{journal}{Z. Phys. B: Condens. Matter} \textbf{\bibinfo{volume}{64}},
  \bibinfo{pages}{189} (\bibinfo{year}{1986});
\bibinfo{author}{\bibfnamefont{Y.}~\bibnamefont{Tokura}},
  \bibinfo{author}{\bibfnamefont{H.}~\bibnamefont{Takagi}}, \bibnamefont{and}
  \bibinfo{author}{\bibfnamefont{S.}~\bibnamefont{Uchida}},
  \bibinfo{journal}{Nature} \textbf{\bibinfo{volume}{337}},
  \bibinfo{pages}{345} (\bibinfo{year}{1989}).

\bibitem[{\citenamefont{Rossat-Mignod et~al.}(1991)\citenamefont{Rossat-Mignod,
  Regnault, Vettier, Bourges, Burlet, Bossy, Henry, and Lapertot}}]{Rossat+91}
\bibinfo{author}{\bibfnamefont{J.}~\bibnamefont{Rossat-Mignod}},
  \bibinfo{author}{\bibfnamefont{L.~P.} \bibnamefont{Regnault}},
  \bibinfo{author}{\bibfnamefont{C.}~\bibnamefont{Vettier}},
  \bibinfo{author}{\bibfnamefont{P.}~\bibnamefont{Bourges}},
  \bibinfo{author}{\bibfnamefont{P.}~\bibnamefont{Burlet}},
  \bibinfo{author}{\bibfnamefont{J.}~\bibnamefont{Bossy}},
  \bibinfo{author}{\bibfnamefont{J.~Y.} \bibnamefont{Henry}}, \bibnamefont{and}
  \bibinfo{author}{\bibfnamefont{G.}~\bibnamefont{Lapertot}},
  \bibinfo{journal}{Physica C} \textbf{\bibinfo{volume}{185}},
  \bibinfo{pages}{86} (\bibinfo{year}{1991});
\bibinfo{author}{\bibfnamefont{H.~F.} \bibnamefont{Fong}},
  \bibinfo{author}{\bibfnamefont{B.}~\bibnamefont{Keimer}},
  \bibinfo{author}{\bibfnamefont{P.~W.} \bibnamefont{Anderson}},
  \bibinfo{author}{\bibfnamefont{D.}~\bibnamefont{Reznik}},
  \bibinfo{author}{\bibfnamefont{F.}~\bibnamefont{Dogan}}, \bibnamefont{and}
  \bibinfo{author}{\bibfnamefont{I.~A.} \bibnamefont{Aksay}},
  \bibinfo{journal}{Phys. Rev. Lett.} \textbf{\bibinfo{volume}{75}},
  \bibinfo{pages}{316} (\bibinfo{year}{1995});
\bibinfo{author}{\bibfnamefont{H.~F.} \bibnamefont{Fong}},
  \bibinfo{author}{\bibfnamefont{P.}~\bibnamefont{Bourges}},
  \bibinfo{author}{\bibfnamefont{Y.}~\bibnamefont{Sidis}},
  \bibinfo{author}{\bibfnamefont{L.~P.} \bibnamefont{Regnault}},
  \bibinfo{author}{\bibfnamefont{A.}~\bibnamefont{Ivanov}},
  \bibinfo{author}{\bibfnamefont{G.~D.} \bibnamefont{Gu}},
  \bibinfo{author}{\bibfnamefont{N.}~\bibnamefont{Koshizuka}},
  \bibnamefont{and} \bibinfo{author}{\bibfnamefont{B.}~\bibnamefont{Keimer}},
  \bibinfo{journal}{Nature} \textbf{\bibinfo{volume}{398}},
  \bibinfo{pages}{588} (\bibinfo{year}{1999}).


\bibitem[{\citenamefont{Dai et~al.}(2001)\citenamefont{Dai, Mook, Hunt, and
  Dogan}}]{Dai+01}
\bibinfo{author}{\bibfnamefont{P.}~\bibnamefont{Dai}},
  \bibinfo{author}{\bibfnamefont{H.~A.} \bibnamefont{Mook}},
  \bibinfo{author}{\bibfnamefont{R.~D.} \bibnamefont{Hunt}}, \bibnamefont{and}
  \bibinfo{author}{\bibfnamefont{F.}~\bibnamefont{Dogan}},
  \bibinfo{journal}{Phys. Rev. B} \textbf{\bibinfo{volume}{63}},
  \bibinfo{pages}{054525} (\bibinfo{year}{2001}).

\bibitem[{\citenamefont{Stock et~al.}(2004)\citenamefont{Stock, Buyers, Liang,
  Peets, Tun, Bonn, Hardy, and Birgeneau}}]{Stock+04}
\bibinfo{author}{\bibfnamefont{C.}~\bibnamefont{Stock}},
  \bibinfo{author}{\bibfnamefont{W.~J.~L.} \bibnamefont{Buyers}},
  \bibinfo{author}{\bibfnamefont{R.}~\bibnamefont{Liang}},
  \bibinfo{author}{\bibfnamefont{D.}~\bibnamefont{Peets}},
  \bibinfo{author}{\bibfnamefont{Z.}~\bibnamefont{Tun}},
  \bibinfo{author}{\bibfnamefont{D.}~\bibnamefont{Bonn}},
  \bibinfo{author}{\bibfnamefont{W.~N.} \bibnamefont{Hardy}}, \bibnamefont{and}
  \bibinfo{author}{\bibfnamefont{R.~J.} \bibnamefont{Birgeneau}},
  \bibinfo{journal}{Phys. Rev. B} \textbf{\bibinfo{volume}{69}},
  \bibinfo{pages}{014502} (\bibinfo{year}{2004}).

\bibitem[{\citenamefont{Hayden et~al.}(2004)\citenamefont{Hayden, Mook, Dai,
  Perring, and Dogan}}]{Hayden+04}
\bibinfo{author}{\bibfnamefont{S.~M.} \bibnamefont{Hayden}},
  \bibinfo{author}{\bibfnamefont{H.~A.} \bibnamefont{Mook}},
  \bibinfo{author}{\bibfnamefont{P.}~\bibnamefont{Dai}},
  \bibinfo{author}{\bibfnamefont{T.~G.} \bibnamefont{Perring}},
  \bibnamefont{and} \bibinfo{author}{\bibfnamefont{F.}~\bibnamefont{Dogan}},
  \bibinfo{journal}{Nature} \textbf{\bibinfo{volume}{429}},
  \bibinfo{pages}{531} (\bibinfo{year}{2004}).

\bibitem[{\citenamefont{Christensen et~al.}(2004)\citenamefont{Christensen,
  McMorrow, Ronnow, Lake, Hayden, Aeppli, Perring, Mangkorntong, Nohara, and
  Takagi}}]{Christensen+04}
\bibinfo{author}{\bibfnamefont{N.~B.} \bibnamefont{Christensen}},
  \bibinfo{author}{\bibfnamefont{D.~F.} \bibnamefont{McMorrow}},
  \bibinfo{author}{\bibfnamefont{H.~M.} \bibnamefont{Ronnow}},
  \bibinfo{author}{\bibfnamefont{B.}~\bibnamefont{Lake}},
  \bibinfo{author}{\bibfnamefont{S.~M.} \bibnamefont{Hayden}},
  \bibinfo{author}{\bibfnamefont{G.}~\bibnamefont{Aeppli}},
  \bibinfo{author}{\bibfnamefont{T.~G.} \bibnamefont{Perring}},
  \bibinfo{author}{\bibfnamefont{M.}~\bibnamefont{Mangkorntong}},
  \bibinfo{author}{\bibfnamefont{M.}~\bibnamefont{Nohara}}, \bibnamefont{and}
  \bibinfo{author}{\bibfnamefont{H.}~\bibnamefont{Tagaki}},
  \bibinfo{journal}{Phys. Rev. Lett.} \textbf{\bibinfo{volume}{93}},
  \bibinfo{pages}{147002} (\bibinfo{year}{2004}).

\bibitem[{\citenamefont{Tranquada et~al.}(2004)\citenamefont{Tranquada, Woo,
  Perring, Goka, Gu, Xu, Fujita, and Yamada}}]{Tranquada+04}
\bibinfo{author}{\bibfnamefont{J.~M.} \bibnamefont{Tranquada}},
  \bibinfo{author}{\bibfnamefont{H.}~\bibnamefont{Woo}},
  \bibinfo{author}{\bibfnamefont{T.~G.} \bibnamefont{Perring}},
  \bibinfo{author}{\bibfnamefont{H.}~\bibnamefont{Goka}},
  \bibinfo{author}{\bibfnamefont{G.~D.} \bibnamefont{Gu}},
  \bibinfo{author}{\bibfnamefont{G.}~\bibnamefont{Xu}},
  \bibinfo{author}{\bibfnamefont{M.}~\bibnamefont{Fujita}}, \bibnamefont{and}
  \bibinfo{author}{\bibfnamefont{K.}~\bibnamefont{Yamada}},
  \bibinfo{journal}{Nature} \textbf{\bibinfo{volume}{429}},
  \bibinfo{pages}{534} (\bibinfo{year}{2004}).

\bibitem[{\citenamefont{Woo et~al.}(2006)\citenamefont{Woo, Dai, Hayden, Mook,
  Dahm, Scalapino, Perring, and Dogan}}]{Woo+06}
\bibinfo{author}{\bibfnamefont{H.}~\bibnamefont{Woo}},
  \bibinfo{author}{\bibfnamefont{P.}~\bibnamefont{Dai}},
  \bibinfo{author}{\bibfnamefont{S.~M.} \bibnamefont{Hayden}},
  \bibinfo{author}{\bibfnamefont{H.~A.} \bibnamefont{Mook}},
  \bibinfo{author}{\bibfnamefont{T.}~\bibnamefont{Dahm}},
  \bibinfo{author}{\bibfnamefont{D.~J.} \bibnamefont{Scalapino}},
  \bibinfo{author}{\bibfnamefont{T.~G.} \bibnamefont{Perring}},
  \bibnamefont{and} \bibinfo{author}{\bibfnamefont{F.}~\bibnamefont{Dogan}},
  \bibinfo{journal}{Nature Physics} \textbf{\bibinfo{volume}{2}},
  \bibinfo{pages}{600} (\bibinfo{year}{2006}).

\bibitem[{\citenamefont{Kivelson et~al.}(2003)\citenamefont{Kivelson, Bindloss,
  Fradkin, Oganesyan, Tranquada, Kapitulnik, and Howald}}]{Kivelson+03}
\bibinfo{author}{\bibfnamefont{S.~A.} \bibnamefont{Kivelson}},
  \bibinfo{author}{\bibfnamefont{I.~P.} \bibnamefont{Bindloss}},
  \bibinfo{author}{\bibfnamefont{E.}~\bibnamefont{Fradkin}},
  \bibinfo{author}{\bibfnamefont{V.}~\bibnamefont{Oganesyan}},
  \bibinfo{author}{\bibfnamefont{J.~M.} \bibnamefont{Tranquada}},
  \bibinfo{author}{\bibfnamefont{A.}~\bibnamefont{Kapitulnik}},
  \bibnamefont{and} \bibinfo{author}{\bibfnamefont{C.}~\bibnamefont{Howald}},
  \bibinfo{journal}{Rev. Mod. Phys.} \textbf{\bibinfo{volume}{75}},
  \bibinfo{pages}{1201} (\bibinfo{year}{2003});
\bibinfo{author}{\bibfnamefont{J.}~\bibnamefont{Zaanen}},
  \bibinfo{author}{\bibfnamefont{O.~Y.} \bibnamefont{Osman}},
  \bibinfo{author}{\bibfnamefont{H.~V.} \bibnamefont{Kruis}},
  \bibinfo{author}{\bibfnamefont{Z.}~\bibnamefont{Nussinov}}, \bibnamefont{and}
  \bibinfo{author}{\bibfnamefont{J.}~\bibnamefont{Tworzydlo}},
  \bibinfo{journal}{Phil. Mag. B} \textbf{\bibinfo{volume}{81}},
  \bibinfo{pages}{1485} (\bibinfo{year}{2001}).

\bibitem[{\citenamefont{Vojta et~al.}(2006)\citenamefont{Vojta, Vojta, and
  Kaul}}]{Vojta+06}
\bibinfo{author}{\bibfnamefont{M.}~\bibnamefont{Vojta}},
  \bibinfo{author}{\bibfnamefont{T.}~\bibnamefont{Vojta}}, \bibnamefont{and}
  \bibinfo{author}{\bibfnamefont{R.~K.} \bibnamefont{Kaul}},
  \bibinfo{journal}{Phys. Rev. Lett.} \textbf{\bibinfo{volume}{97}},
  \bibinfo{pages}{097001} (\bibinfo{year}{2006}).

\bibitem[{\citenamefont{Demler and Zhang}(1998)}]{Demler+98}
\bibinfo{author}{\bibfnamefont{E.}~\bibnamefont{Demler}} \bibnamefont{and}
  \bibinfo{author}{\bibfnamefont{S.-C.} \bibnamefont{Zhang}},
  \bibinfo{journal}{Nature} \textbf{\bibinfo{volume}{396}},
  \bibinfo{pages}{733} (\bibinfo{year}{1998});
\bibinfo{author}{\bibfnamefont{S.}~\bibnamefont{Sachdev}},
  \bibinfo{journal}{Rev. Mod. Phys.} \textbf{\bibinfo{volume}{75}},
  \bibinfo{pages}{913} (\bibinfo{year}{2003}).

\bibitem[{\citenamefont{Norman}(2000)}]{Norman00}
\bibinfo{author}{\bibfnamefont{M.~R.} \bibnamefont{Norman}},
  \bibinfo{journal}{Phys. Rev. B} \textbf{\bibinfo{volume}{61}},
  \bibinfo{pages}{14751} (\bibinfo{year}{2000}).

\bibitem[{\citenamefont{Norman}(2001)}]{Norman01}
\bibinfo{author}{\bibfnamefont{M.~R.} \bibnamefont{Norman}},
  \bibinfo{journal}{Phys. Rev. B} \textbf{\bibinfo{volume}{63}},
  \bibinfo{pages}{092509} (\bibinfo{year}{2001});
\bibinfo{author}{\bibfnamefont{D.}~\bibnamefont{Manske}},
  \bibinfo{author}{\bibfnamefont{I.}~\bibnamefont{Eremin}}, \bibnamefont{and}
  \bibinfo{author}{\bibfnamefont{K.~H.} \bibnamefont{Bennemann}},
  \bibinfo{journal}{Phys. Rev. B} \textbf{\bibinfo{volume}{63}},
  \bibinfo{pages}{054517} (\bibinfo{year}{2001}).

\bibitem[{\citenamefont{Eremin et~al.}(2005)\citenamefont{Eremin, Morr,
  Chubukov, Bennemann, and Norman}}]{Eremin+05}
\bibinfo{author}{\bibfnamefont{I.}~\bibnamefont{Eremin}},
  \bibinfo{author}{\bibfnamefont{D.~K.} \bibnamefont{Morr}},
  \bibinfo{author}{\bibfnamefont{A.~V.} \bibnamefont{Chubukov}},
  \bibinfo{author}{\bibfnamefont{K.~H.} \bibnamefont{Bennemann}},
  \bibnamefont{and} \bibinfo{author}{\bibfnamefont{M.~R.}
  \bibnamefont{Norman}}, \bibinfo{journal}{Phys. Rev. Lett.}
  \textbf{\bibinfo{volume}{94}}, \bibinfo{pages}{147001}
  (\bibinfo{year}{2005}).

\bibitem[{\citenamefont{Eschrig}(2006)}]{Eschrig06}
\bibinfo{author}{\bibfnamefont{M.}~\bibnamefont{Eschrig}},
  \bibinfo{journal}{Adv. in Phys.} \textbf{\bibinfo{volume}{55}},
  \bibinfo{pages}{47} (\bibinfo{year}{2006}).

\bibitem[{\citenamefont{Wilson et~al.}(2006{\natexlab{a}})\citenamefont{Wilson,
  Dai, Li, Chi, Kang, and Lynn}}]{Wilson+06a}
\bibinfo{author}{\bibfnamefont{S.~D.} \bibnamefont{Wilson}},
  \bibinfo{author}{\bibfnamefont{P.}~\bibnamefont{Dai}},
  \bibinfo{author}{\bibfnamefont{S.}~\bibnamefont{Li}},
  \bibinfo{author}{\bibfnamefont{S.}~\bibnamefont{Chi}},
  \bibinfo{author}{\bibfnamefont{H.~J.} \bibnamefont{Kang}}, \bibnamefont{and}
  \bibinfo{author}{\bibfnamefont{J.~W.} \bibnamefont{Lynn}},
  \bibinfo{journal}{Nature} \textbf{\bibinfo{volume}{442}}, \bibinfo{pages}{59}
  (\bibinfo{year}{2006}{\natexlab{a}}).

\bibitem[{\citenamefont{Bulut and Scalapino}(1996)}]{Bulut+96}
\bibinfo{author}{\bibfnamefont{N.}~\bibnamefont{Bulut}} \bibnamefont{and}
  \bibinfo{author}{\bibfnamefont{D.~J.} \bibnamefont{Scalapino}},
  \bibinfo{journal}{Phys. Rev. B} \textbf{\bibinfo{volume}{53}},
  \bibinfo{pages}{5149} (\bibinfo{year}{1996}).

\bibitem[{\citenamefont{Das et~al.}(2006)\citenamefont{Das, Markiewicz, and
  Bansil}}]{Das+06}
\bibinfo{author}{\bibfnamefont{T.}~\bibnamefont{Das}},
  \bibinfo{author}{\bibfnamefont{R.~S.} \bibnamefont{Markiewicz}},
  \bibnamefont{and} \bibinfo{author}{\bibfnamefont{A.}~\bibnamefont{Bansil}},
  \bibinfo{journal}{Phys. Rev. B} \textbf{\bibinfo{volume}{74}},
  \bibinfo{pages}{020506(R)} (\bibinfo{year}{2006}).

\bibitem[{\citenamefont{Matsui et~al.}(2005)\citenamefont{Matsui, Terashima,
  Sato, Takahashi, Fujita, and Yamada}}]{Matsui+05}
\bibinfo{author}{\bibfnamefont{H.}~\bibnamefont{Matsui}},
  \bibinfo{author}{\bibfnamefont{K.}~\bibnamefont{Terashima}},
  \bibinfo{author}{\bibfnamefont{T.}~\bibnamefont{Sato}},
  \bibinfo{author}{\bibfnamefont{T.}~\bibnamefont{Takahashi}},
  \bibinfo{author}{\bibfnamefont{M.}~\bibnamefont{Fujita}}, \bibnamefont{and}
  \bibinfo{author}{\bibfnamefont{K.}~\bibnamefont{Yamada}},
  \bibinfo{journal}{Phys. Rev. Lett.} \textbf{\bibinfo{volume}{95}},
  \bibinfo{pages}{017003} (\bibinfo{year}{2005}).

\bibitem[{\citenamefont{Shan et~al.}(2005)\citenamefont{Shan, Huang, Gao, Wang,
  Li, Dai, Zhou, Xiong, Ti, and Wen}}]{Shan+05}
\bibinfo{author}{\bibfnamefont{L.}~\bibnamefont{Shan}},
  \bibinfo{author}{\bibfnamefont{Y.}~\bibnamefont{Huang}},
  \bibinfo{author}{\bibfnamefont{H.}~\bibnamefont{Gao}},
  \bibinfo{author}{\bibfnamefont{Y.}~\bibnamefont{Wang}},
  \bibinfo{author}{\bibfnamefont{S.~L.} \bibnamefont{Li}},
  \bibinfo{author}{\bibfnamefont{P.~C.} \bibnamefont{Dai}},
  \bibinfo{author}{\bibfnamefont{F.}~\bibnamefont{Zhou}},
  \bibinfo{author}{\bibfnamefont{J.~W.} \bibnamefont{Xiong}},
  \bibinfo{author}{\bibfnamefont{W.~X.} \bibnamefont{Ti}}, \bibnamefont{and}
  \bibinfo{author}{\bibfnamefont{H.~H.} \bibnamefont{Wen}},
  \bibinfo{journal}{Phys. Rev. B} \textbf{\bibinfo{volume}{72}},
  \bibinfo{pages}{144506} (\bibinfo{year}{2005}).

\bibitem[{\citenamefont{Blonder et~al.}(1982)\citenamefont{Blonder, Tinkham,
  and Klapwijk}}]{Blonder+82}
\bibinfo{author}{\bibfnamefont{G.~E.} \bibnamefont{Blonder}},
  \bibinfo{author}{\bibfnamefont{M.}~\bibnamefont{Tinkham}}, \bibnamefont{and}
  \bibinfo{author}{\bibfnamefont{T.~M.} \bibnamefont{Klapwijk}},
  \bibinfo{journal}{Phys. Rev. B} \textbf{\bibinfo{volume}{25}},
  \bibinfo{pages}{4515} (\bibinfo{year}{1982});
\bibinfo{author}{\bibfnamefont{Y.}~\bibnamefont{Tanaka}} \bibnamefont{and}
  \bibinfo{author}{\bibfnamefont{S.}~\bibnamefont{Kashiwaya}},
  \bibinfo{journal}{Phys. Rev. B} \textbf{\bibinfo{volume}{53}},
  \bibinfo{pages}{9371} (\bibinfo{year}{1996}).

\bibitem[{\citenamefont{Reznik et~al.}()\citenamefont{Reznik, Ismer, Eremin,
  Pintschovius, M.Arai, Endoh, Masui, and Tajima}}]{Reznik+06}
\bibinfo{author}{\bibfnamefont{D.}~\bibnamefont{Reznik}},
  \bibinfo{author}{\bibfnamefont{J.~P.} \bibnamefont{Ismer}},
  \bibinfo{author}{\bibfnamefont{I.}~\bibnamefont{Eremin}},
  \bibinfo{author}{\bibfnamefont{L.}~\bibnamefont{Pintschovius}},
  \bibinfo{author}{\bibnamefont{M.Arai}},
  \bibinfo{author}{\bibfnamefont{Y.}~\bibnamefont{Endoh}},
  \bibinfo{author}{\bibfnamefont{T.}~\bibnamefont{Masui}}, \bibnamefont{and}
  \bibinfo{author}{\bibfnamefont{S.}~\bibnamefont{Tajima}},
  \bibinfo{note}{\coma /0610755}.

\bibitem[{\citenamefont{Shirane et~al.}(2002)\citenamefont{Shirane, Shapiro,
  and Tranquada}}]{Shirane+02}
\bibinfo{author}{\bibfnamefont{G.}~\bibnamefont{Shirane}},
  \bibinfo{author}{\bibfnamefont{S.}~\bibnamefont{Shapiro}}, \bibnamefont{and}
  \bibinfo{author}{\bibfnamefont{J.~M.} \bibnamefont{Tranquada}},
  \emph{\bibinfo{title}{Neutron Scattering with a Triple Axis Spectrometer}}
  (\bibinfo{publisher}{Cambridge University Press},
  \bibinfo{address}{Cambridge, England}, \bibinfo{year}{2002}).

\bibitem[{\citenamefont{Wilson et~al.}(2006{\natexlab{b}})\citenamefont{Wilson,
  Li, Woo, Dai, Mook, Frost, Komiya, , and Ando}}]{Wilson+06b}
\bibinfo{author}{\bibfnamefont{S.~D.} \bibnamefont{Wilson}},
  \bibinfo{author}{\bibfnamefont{S.}~\bibnamefont{Li}},
  \bibinfo{author}{\bibfnamefont{H.}~\bibnamefont{Woo}},
  \bibinfo{author}{\bibfnamefont{P.}~\bibnamefont{Dai}},
  \bibinfo{author}{\bibfnamefont{H.~A.} \bibnamefont{Mook}},
  \bibinfo{author}{\bibfnamefont{C.~D.} \bibnamefont{Frost}},
  \bibinfo{author}{\bibfnamefont{S.}~\bibnamefont{Komiya}}, , \bibnamefont{and}
  \bibinfo{author}{\bibfnamefont{Y.}~\bibnamefont{Ando}},
  \bibinfo{journal}{Phys. Rev. Lett.} \textbf{\bibinfo{volume}{96}},
  \bibinfo{pages}{157001} (\bibinfo{year}{2006}{\natexlab{b}}).

\bibitem[{\citenamefont{Fujita et~al.}(2006)\citenamefont{Fujita, Matsuda, Fak,
  Frost, and Yamada}}]{Fujita+06}
\bibinfo{author}{\bibfnamefont{M.}~\bibnamefont{Fujita}},
  \bibinfo{author}{\bibfnamefont{M.}~\bibnamefont{Matsuda}},
  \bibinfo{author}{\bibfnamefont{B.}~\bibnamefont{Fak}},
  \bibinfo{author}{\bibfnamefont{C.~D.} \bibnamefont{Frost}}, \bibnamefont{and}
  \bibinfo{author}{\bibfnamefont{K.}~\bibnamefont{Yamada}},
  \bibinfo{journal}{J. Phys. Soc. Jpn.} \textbf{\bibinfo{volume}{75}},
  \bibinfo{pages}{093704} (\bibinfo{year}{2006}).



\bibitem[{\citenamefont{Ismer et~al.}()\citenamefont{Ismer, Eremin, Rossi, and
  Morr}}]{Ismer+07}
\bibinfo{author}{\bibfnamefont{J.~P.} \bibnamefont{Ismer}},
  \bibinfo{author}{\bibfnamefont{I.}~\bibnamefont{Eremin}},
  \bibinfo{author}{\bibfnamefont{E.}~\bibnamefont{Rossi}}, \bibnamefont{and}
  \bibinfo{author}{\bibfnamefont{D.~K.} \bibnamefont{Morr}},
  \bibinfo{note}{\coma /0702375}.

\bibitem[{\citenamefont{Bourges et~al.}(1997)\citenamefont{Bourges, Casalta,
  Ivanov, and Petitgrand}}]{Bourges+97}
\bibinfo{author}{\bibfnamefont{P.}~\bibnamefont{Bourges}},
  \bibinfo{author}{\bibfnamefont{H.}~\bibnamefont{Casalta}},
  \bibinfo{author}{\bibfnamefont{A.~S.} \bibnamefont{Ivanov}},
  \bibnamefont{and}
  \bibinfo{author}{\bibfnamefont{D.}~\bibnamefont{Petitgrand}},
  \bibinfo{journal}{Phys. Rev. Lett.} \textbf{\bibinfo{volume}{79}},
  \bibinfo{pages}{4906} (\bibinfo{year}{1997}).

\bibitem[{\citenamefont{Tranquada et~al.}(2006)\citenamefont{Tranquada, Woo,
  Perring, Goka, Gu, Xu, Fujita, and Yamada}}]{Tranquada+06}
\bibinfo{author}{\bibfnamefont{J.~M.} \bibnamefont{Tranquada}},
  \bibinfo{author}{\bibfnamefont{H.}~\bibnamefont{Woo}},
  \bibinfo{author}{\bibfnamefont{T.~G.} \bibnamefont{Perring}},
  \bibinfo{author}{\bibfnamefont{H.}~\bibnamefont{Goka}},
  \bibinfo{author}{\bibfnamefont{G.~D.} \bibnamefont{Gu}},
  \bibinfo{author}{\bibfnamefont{G.}~\bibnamefont{Xu}},
  \bibinfo{author}{\bibfnamefont{M.}~\bibnamefont{Fujita}}, \bibnamefont{and}
  \bibinfo{author}{\bibfnamefont{K.}~\bibnamefont{Yamada}},
  \bibinfo{journal}{Journal of Physics and Chemistry of Solids}
  \textbf{\bibinfo{volume}{67}}, \bibinfo{pages}{511} (\bibinfo{year}{2006}).

\bibitem[{\citenamefont{Kr\"uger and Scheidl}(2003)}]{Kruger+03}
\bibinfo{author}{\bibfnamefont{F.}~\bibnamefont{Kr\"uger}} \bibnamefont{and}
  \bibinfo{author}{\bibfnamefont{S.}~\bibnamefont{Scheidl}},
  \bibinfo{journal}{Phys. Rev. B} \textbf{\bibinfo{volume}{67}},
  \bibinfo{pages}{134512} (\bibinfo{year}{2003});
\bibinfo{author}{\bibfnamefont{F.}~\bibnamefont{Kr\"uger}} \bibnamefont{and}
  \bibinfo{author}{\bibfnamefont{S.}~\bibnamefont{Scheidl}},
  \bibinfo{journal}{Phys. Rev. B} \textbf{\bibinfo{volume}{70}},
  \bibinfo{pages}{064421} (\bibinfo{year}{2004});
\bibinfo{author}{\bibfnamefont{M.}~\bibnamefont{Vojta}} \bibnamefont{and}
  \bibinfo{author}{\bibfnamefont{T.}~\bibnamefont{Ulbricht}},
  \bibinfo{journal}{Phys. Rev. Lett.} \textbf{\bibinfo{volume}{93}},
  \bibinfo{pages}{127002} (\bibinfo{year}{2004});
\bibinfo{author}{\bibfnamefont{G.~S.} \bibnamefont{Uhrig}},
  \bibinfo{author}{\bibfnamefont{K.~P.} \bibnamefont{Schmidt}},
  \bibnamefont{and}
  \bibinfo{author}{\bibfnamefont{M.}~\bibnamefont{Gruninger}},
  \bibinfo{journal}{Phys. Rev. Lett.} \textbf{\bibinfo{volume}{93}},
  \bibinfo{pages}{267003} (\bibinfo{year}{2004});
\bibinfo{author}{\bibfnamefont{B.~M.} \bibnamefont{Andersen}} \bibnamefont{and}
  \bibinfo{author}{\bibfnamefont{P.}~\bibnamefont{Hedegard}},
  \bibinfo{journal}{Phys. Rev. Lett.} \textbf{\bibinfo{volume}{95}},
  \bibinfo{pages}{037002} (\bibinfo{year}{2005});
\bibinfo{author}{\bibfnamefont{G.}~\bibnamefont{Seibold}} \bibnamefont{and}
  \bibinfo{author}{\bibfnamefont{J.}~\bibnamefont{Lorenzana}},
  \bibinfo{journal}{Phys. Rev. Lett.} \textbf{\bibinfo{volume}{94}},
  \bibinfo{pages}{107006} (\bibinfo{year}{2005});
\bibinfo{author}{\bibfnamefont{G.}~\bibnamefont{Seibold}} \bibnamefont{and}
  \bibinfo{author}{\bibfnamefont{J.}~\bibnamefont{Lorenzana}},
  \bibinfo{journal}{Phys. Rev. B} \textbf{\bibinfo{volume}{73}},
  \bibinfo{pages}{144515} (\bibinfo{year}{2006});
\bibinfo{author}{\bibfnamefont{D.~X.} \bibnamefont{Yao}},
  \bibinfo{author}{\bibfnamefont{E.~W.} \bibnamefont{Carlson}},
  \bibnamefont{and} \bibinfo{author}{\bibfnamefont{D.~K.}
  \bibnamefont{Campbell}}, \bibinfo{journal}{Phys. Rev. Lett.}
  \textbf{\bibinfo{volume}{97}}, \bibinfo{pages}{017003}
  (\bibinfo{year}{2006}).

\end{thebibliography}
\end{document}